\documentstyle [12pt]{ article}
\begin{document}
\thispagestyle{empty}
\begin{flushright}
LPC 93 06\\
hep-ph/9501225
\end{flushright}
\begin{center}
{\bf GLUEBALL PLUS PION PRODUCTION}\\[5pt]
{\bf IN HARD EXCLUSIVE TWO-PHOTON PROCESSES}\\[20pt]
{\bf M. A. ICHOLA and J. PARISI}\\[20pt]
{\bf Laboratoire de Physique Corpusculaire, Coll\`{e}ge de
France}\\[5pt]
{\bf IN2P3-CNRS} \\ [5 pt]
{\bf 11, Place Marcelin Berthelot, F-75231 Paris Cedex 05, France}
\end{center}
\vspace*{6cm}
\begin{center}
{\bf Abstract} \\
\end{center}
\hspace*{1cm} We here compute the reaction
$ \gamma \; \gamma \rightarrow G \; \pi^{0} $ for
various glueball candidates $ G $ and their assumed quantum states,
using a
non-relativistic gluon bound-state model for the glueball. \\[30 pt]
\newpage
\section{Introduction}
\hspace*{1cm} A model for computing the production and decay of
glueballs
($ G $) made up of
two gluons, in any quantum state, has been proposed by Kada et al.
{\bf [1]},
who used it in order to calculate the processes
$ J/\Psi \rightarrow G \;  \gamma $ and
$ G \Leftrightarrow \gamma \; \gamma $. That model was later
generalized
for more complex reactions by Houra-Yaou et al. {\bf [2]} ; the
generalized
formalism was applied, in particular, to the calculation of glueball
production in high-energy hadron collisions. \\[10 pt]
\hspace*{1cm} The values obtained in Ref. {\bf [1]} for
$ \Gamma(G \rightarrow \gamma \;  \gamma) $,
 considering the main existing glueball candidates, are rather small,
and
systematically below the experimental limits established till now in
measurements of $ \gamma \; \gamma \rightarrow G$ . The purpose of
this
paper is to compute an alternative photon-photon collision process
producing
glueballs, i.e. $ \gamma \; \gamma \rightarrow G \; \pi^{0}$. The
model used in this study is again that of Ref. {\bf [2]}.\\[10pt]
\hspace*{1cm} Let us notice that the pseudoscalar glueball candidate
$\eta(1440)$ was seen in
various measurements of radiative $J/\Psi$ decays {\bf [3]}. On the
other hand
, two new analyses of $J/\Psi \rightarrow \eta (1440) \gamma$
performed by the
Mark III and the DM2 Collaboration {\bf [4]} both seem to show that,
instead of
corresponding to a single pseudoscalar resonance, the $\eta (1440)$
peak should
be due to three different structures (two pseudoscalar and one axial)
located
close to each other in the 1400-1500 MeV range. However the
conclusions of the
two analyses are contradictory.\\[10pt]
\hspace*{1cm} As for the tensor glueball candidate $f_{2}(1720)$, it
was seen
as well by
various experimental groups in radiative $J/\Psi$ decay {\bf [5]} and
also by
the WA 76 Collaboration in double-diffractive pp collisions {\bf
[6]}. Here
again some doubt has been cast on the true nature of this resonance
by a
recent analysis of Mark III {\bf [7]}. According to this analysis it
should
rather be a scalar particle, $f_{0}(1710)$, and is unlikely to be a
glueball.
Thus there is, here again, an obvious disagreement between various
experimental groups (see {\bf [8]}).\\[10pt]
\hspace*{1cm} Finally, as regards the $X(2220)$, it was also found by
the Mark
III
Collaboration in radiative $J/\Psi$ decay, and was assigned the
quantum
numbers : $J$ even (without further precision), $P = +$ ~{\bf [9]}.
On the
other hand, it was recently seen by two experimental groups in
hadronic
reactions, but they disagreed on its spin value ($J^{P} = 2^{+}$ vs.
$J^{P} = 4^{+})$ {\bf [10]} ; specialists now tend to
believe that the $X$ is rather a $s \bar{s}$ quarkonium state
belonging to the
nonet $^{3}F$.\\[10pt]
\hspace*{1cm} It is obvious that, in all three cases considered, no
firm
conclusion has yet
been reached as to the nature of these resonances. Therefore we
consider that
none of them has been discarded as a glueball candidate, and we
continue
treating them as in previous papers {\bf [1, 2]}.
\section{Details of calculation}
\hspace*{1cm} Assuming, here again, the glueball to be a weakly bound
state of
two non-relativistic gluons, we are led (see Sec. II of Ref. {\bf
[2])} to the
formula
\begin{eqnarray}
{\cal M}^{\lambda \; \lambda', \; \Lambda}_{\gamma \; \gamma'
\rightarrow G \; \pi^{0}}(s, \Theta) &=& f_{L} \; \lim_{\beta
\rightarrow 0}
\frac{1}{\beta^{L}} \int  \frac{d(\cos \theta) \; d \phi}{4 \; \pi}
\times
\nonumber \\[15 pt]
& & \times \sum_{\lambda_{1}, \; \lambda_{2}}
\zeta^{L \;S \;J \; \Lambda}_{\lambda_{1} \;
\lambda_{2}}(\theta,\phi) \;
{\cal M}^{\lambda \; \lambda', \; \lambda_{1} \; \lambda_{2}}_{
\gamma \; \gamma' \rightarrow g_{1} \; g_{2} \; \pi^{0}}
(s, \Theta, \theta, \phi)
\end{eqnarray}
connecting the helicity amplitudes of the process
$ \gamma \; \gamma' \rightarrow G \; \pi^{0}$ with those of the
subprocess
$ \gamma \; \gamma' \rightarrow g_{1} \; g_{2} \; \pi^{0}$. In
formula (1),
$ s $ and $ \Theta $ are, respectively, the total energy squared and
the pion
emission angle in the $ \gamma \; \gamma' $ c.m. frame, while
$ \theta \; (\phi) $
is the orbital (azimuthal) emission angle of either gluon in their
c.m. frame
(the glueball rest frame) ; see Fig. 1. We call J, L, S,
respectively,
the total spin, the orbital angular momentum and the intrinsic spin
of the
glueball, while $ \Lambda $ is its spin component along the z axis of
Fig. 1.
In addition we call $ \lambda, \; \lambda'$ the helicities of the
photons
$ \gamma, \; \gamma' $, while $ \lambda_{1}, \; \lambda_{2}$ are
those of the
gluons $ g_{1}, \; g_{2} $ (all helicities being defined in the
glueball rest
frame). The angular projection function $ \zeta^{LSJ
\Lambda}_{\lambda_{1}
 \; \lambda_{2}}(\theta, \phi) $ is defined as :
\begin{eqnarray}
\zeta^{L \;S \; J \; \Lambda}_{\lambda_{1} \; \lambda_{2}}(\theta,
\phi) &=&
d^{J}_{\Lambda \bar{\Lambda}} \; e^{-i \; \Lambda \; \phi} \;
\langle L\;S\;0\;\bar{\Lambda}\;|\;L\;S\;J\;\bar{\Lambda}\rangle
\langle
1\;1\;\lambda_{1}\;-\lambda_{2}\;|\;1\;1\;S\;\bar{\Lambda}\rangle
\end{eqnarray}
where $ \bar{\Lambda} = \lambda_{1}-\lambda_{2} $. Finally, $ \beta $
is the
velocity of either gluon in the glueball rest frame, while $ f_{L} $
is given
by
\begin{eqnarray}
f_{L} &=& \sqrt{\frac{2 \; L \; + \; 1}{2 \; \pi \; M}} \;
\left(- \frac{2 \; i}{M}\right)^{L}\;\frac{(2\;L\;+\;1)!!}{L!} \;
\left[\left(\frac{d}{dr} \right)^{L}
R_{L}(r) \right]_{r \rightarrow 0}
\end{eqnarray}
where $ M $ is the glueball mass, and $ R_{L}(r) $ its radial wave
function
in configuration space. \\[10 pt]
\hspace*{1cm} As in Ref. {\bf [2]}, we assume the glueball to be
relativistic in the
$ \gamma \; \gamma' $ c.m. frame : $ M/\sqrt{s} \rightarrow 0 $. In
that
approximation, the gluons $ g_{1}, \; g_{2} $ are also treated as
massless in the subprocess
$ \gamma \; \gamma' \rightarrow g_{1} \; g_{2} \;\pi^{0} $. The
latter
subprocess involves 16 helicity amplitudes. Noticing that, due
to angular momentum and parity conservation, those amplitudes remain
unchanged
when all photon and gluon helicities are reversed, only eight of them
are
needed.\\[10 pt]
\hspace*{1cm} For the calculation of those amplitudes, we use the
Brodsky-Lepage model {\bf [11]}, i.e. :
\begin{eqnarray}
{\cal M}^{\lambda \; \lambda', \; \lambda_{1} \lambda_{2}}_{
\gamma \; \gamma' \rightarrow g_{1} \; g_{2} \; \pi^{0}}
(s, \Theta, \theta, \phi) &=& \int \; dx \; \Phi_{\pi}(x) \;
{\cal M}^{\lambda \; \lambda', \; \lambda_{1} \lambda_{2}}_{
\gamma \; \gamma' \rightarrow g_{1} \; g_{2} \; (q \; \bar{q})_{PS}}
(x, s, \Theta, \theta, \phi)
\end{eqnarray}
where $ \Phi_{\pi} $ is the pion distribution amplitude , while in
the
calculation of \\
$ {\cal M}^{\lambda \; \lambda', \; \lambda_{1} \lambda_{2}}_{
\gamma \; \gamma' \rightarrow g_{1} \; g_{2} \; (q \; \bar{q})_{PS}}
$ one
makes the substitution :
$ v_{\bar{q}} \bar{u}_{q} \rightarrow \gamma_{5} p_{\pi}/\sqrt{2} $,
taking into account the fact that the $ (q \; \bar{q}) $ system is in
a
pseudoscalar state. The helicity amplitudes
$ {\cal M}^{\lambda \; \lambda', \; \lambda_{1} \lambda_{2}}_{
\gamma \; \gamma' \rightarrow g_{1} \; g_{2} \; (q \; \bar{q})_{PS}}
$ needed
are obtained, at lowest order in QCD, by summing over the
contributions of all
diagrams of Fig. 2.\\[10 pt]
\hspace*{1cm} We then remain with the task of applying formulas (4)
and (1).
Actually, we find more convenient to integrate over $ \theta, \phi $
first,
applying formula (1), and to leave the convolution over $
\phi_{\pi}(x) $ for
the following stage of our calculation. \\[10 pt]
\hspace*{1cm} Formula (1), where we substitute
$ (q \; \bar{q})_{PS} $ for $ \pi^{0}$, leads us,
for the various quantum states considered (for the main glueball
candidates) to
the expressions
$ {\cal M}^{\lambda\;\lambda',\;\Lambda}_{\gamma\;\gamma'\rightarrow
G\;
(q \; \bar{q})_{PS}} $shown hereafter. We here fixed $ \lambda = +
1$, since
amplitudes with $ \lambda = - 1$ are derived therefrom by applying
the relation
${\cal M}^{-\lambda \; -\lambda', \; -\Lambda}= (-1)^{J+L-\Lambda -1}
{\cal M}^{\lambda \; \lambda', \; \Lambda} $. \\
\hspace*{1cm}All other helicity amplitudes are vanishing at that
order.
\\[10 pt]
\hspace*{1cm}Let $ \mbox{cos} \Theta=u $,~~~~ $ N=x\;(1-x)\;s $,~~~~
$ y=-2\;(1-x)\;(1+u) $,\\
\hspace*{1cm}$ y_{1}=1+u\;(1-2\;x) $~~~~ and~~~~  $ z=1+u-2\;x $.
\\[15 pt]
\hspace*{1cm}In addition we set :
\begin{eqnarray*}
A_{1} & = & -2(1-u)(1-2x)-u\frac{y}{z} \\[10pt]
B_{1} & = & -\frac{2u}{z}\\[10pt]
C_{1} & = & 4x^{2}(1+u)+2(u+3)(1-2x)+2(1-x)(1-u^{2})
+(4(x-1)^{2}-1-u^{2})\frac{y}{z} \\[10pt]
D_{1} & = & -4(1-x)(1+u)\\[10pt]
A_{2} & = & -4(1-x)\frac{y}{z}\\[10pt]
B_{2} & = & 4\frac{(1-x)}{z}\\[10pt]
C_{2} & = & 2(1+u)(1-x)(-2x+1-u)
+(4(x-1)^{2}+1-u^{2})\frac{y}{z} \\[10pt]
D_{2} & = & 4(1-x(1-u))
\end{eqnarray*}
\hspace*{1cm}We get :
\begin{eqnarray}
{\bf J=0, \; L=S=0} & & \nonumber \\
{\cal M}^{+ \; +, \; 0}_{\gamma \; \gamma' \rightarrow G \;
(q \; \bar{q})_{PS}} &=& \; \; \; \; -\; 8 \;
\pi\frac{B_{1}\;y_{1}-A_{1}}
{y_{1}\;N} \\[15pt]
{\cal M}^{+ \; -, \; 0}_{\gamma \; \gamma' \rightarrow G \;
(q \; \bar{q})_{PS}} &=& \; \; \; \;- \; 8 \;
\pi\frac{B_{2}\;y_{1}-A_{2}}
{y_{1}\;N}\\[25pt]
{\bf J=0, \; L=S=1} & & \nonumber \\
{\cal M}^{+ \; +, \; 0}_{\gamma \; \gamma' \rightarrow G \;
(q \; \bar{q})_{PS}} &=& \; \; \; \; \; \; \; \; 8 \; \pi \;
\frac{C_{1}\;z}
{3 \; y_{1}^{2}\;N } \\[15pt]
{\cal M}^{+ \; -, \; 0}_{\gamma \; \gamma' \rightarrow G \;
(q \; \bar{q})_{PS}} &=& \; \; \; \; - \; 8 \; \pi \;\frac{C_{2}\;z}
{3y_{1}^{2}\; N }  \\[25pt]
{\bf J=2, \; L=0, \; S=2} & & \nonumber \\
{\cal M}^{+ \; +, \; 0}_{\gamma \; \gamma' \rightarrow G \;
(q \; \bar{q})_{PS}} &=& \; \; \; - \; 16 \; \pi \;
\frac{B_{1}\;y_{1}-A_{1}}{y_{1}\;N}\\[15 pt]
{\cal M}^{+ \; +, \; 2}_{\gamma \; \gamma' \rightarrow G \;
(q \; \bar{q})_{PS}} &=& \;\;\; - \; 16 \; \pi \; \frac{D_{1}}
{y_{1} \; N} \\[15 pt]
{\cal M}^{+ \; -, \; 0}_{\gamma \; \gamma' \rightarrow G \;
(q \; \bar{q})_{PS}} &=& \;\;\;- \; 16 \; \pi
\frac{B_{2}\;y_{1}-A_{2}}{y_{1}\;N} \\[15 pt]
{\cal M}^{+ \; -, - 2}_{\gamma \; \gamma' \rightarrow G \;
(q \; \bar{q})_{PS}} &=& \;\;\;- \; 16 \; \pi \; \frac{D_{2}}
{y_{1} \; N} \\[25pt]
{\bf J=2, \; L=2, \; S=0} & & \nonumber \\
{\cal M}^{+ \; +, \; 0}_{\gamma \; \gamma' \rightarrow G \;
(q \; \bar{q})_{PS}} &=& \; \; \; \; \; \; 32 \; \pi \;
\frac{B_{1}\;y_{1}^{3}-A_{1}\;(z^{2}+y_{1}^{2})}{15\;y_{1}^{3}\;
N}\\[15 pt]
{\cal M}^{+ \; +, \; 2}_{\gamma \; \gamma' \rightarrow G \;
(q \; \bar{q})_{PS}} &=& \;\;\;- 64 \; \pi \; \frac{D_{1}}
{15 \;y_{1} \; N}\\[15 pt]
{\cal M}^{+ \; -, \; 0}_{\gamma \; \gamma' \rightarrow G \;
(q \; \bar{q})_{PS}} &=& \; \; \; \; \; \; 32 \; \pi \;
\frac{B_{2}\;y_{1}^{3}-A_{2}\;(z^{2}+y_{1}^{2})}{15\;y_{1}^{3}\;
N}\\[15 pt]
{\cal M}^{+ \; -, \; -2}_{\gamma \; \gamma' \rightarrow G \;
(q \; \bar{q})_{PS}} &=& \; \; \;-64 \; \pi \; \frac{D_{2}}
{15 \;y_{1} \; N} \\[25pt]
{\bf J=2, \; L=S=2} & & \nonumber \\
{\cal M}^{+ \; +, \; 0}_{\gamma \; \gamma' \rightarrow G \;
(q \; \bar{q})_{PS}} &=& \; \; \; \; \;8 \; \pi \;
\frac{17\;B_{1}\;y_{1}^{3}+A_{1}\;(4z^{2}-17y_{1}^{2})}
{15\;y_{1}^{3}\;N} \\[15 pt]
{\cal M}^{+ \; +, \; 2}_{\gamma \; \gamma' \rightarrow G \;
(q \; \bar{q})_{PS}} &=& \; \; \; 16 \; \pi \;
\frac{D_{1}\; (4z^{2}-3y_{1}^{2})}{15 \; y_{1}^{3}\;N}\\[15 pt]
{\cal M}^{+ \; -, \; 0}_{\gamma \; \gamma' \rightarrow G \;
(q \; \bar{q})_{PS}} &=& \; \; \; \;8 \; \pi \;
\frac{17\;B_{2}\; y_{1}^{3}+A_{2}\;(4z^{2}-17y_{1}^{2})}
{15 \; y_{1}^{3}\; N}\\[15 pt]
{\cal M}^{+ \; -, \; -2}_{\gamma \; \gamma' \rightarrow G \;
(q \; \bar{q})_{PS}} &=& \; \; \; 16 \; \pi \;
\frac{D_{1}\;(4z^{2}-3y_{1}^{2})}{15 \; y_{1}\;N} \\[25pt]
{\bf J=4, \; L=S=2} & & \nonumber \\
{\cal M}^{+ \; +, \; 0}_{\gamma \; \gamma' \rightarrow G \;
(q \; \bar{q})_{PS}} &=& \; \;  128 \; \pi
\frac{B_{1}\;y_{1}^{3}-A_{1}(z^{2}+y_{1}^{2})}{75 \; y_{1}^{3} \;
N}\\[15 pt]
{\cal M}^{+ \; +, \; 2}_{\gamma \; \gamma' \rightarrow G \;
(q \; \bar{q})_{PS}} &=& \; \;  \; 64 \; \pi \;
\frac{D_{1}\;(z^{2}+y_{1}^{2})}{15 \; y_{n1}^{3} \; N}\\[15 pt]
{\cal M}^{+ \; -, \; 0}_{\gamma \; \gamma' \rightarrow G \;
(q \; \bar{q})_{PS}} &=& \; \;  128\; \pi \;
\frac{B_{2}\;y_{1}^{3}-A_{2}\;(z^{2}+y_{1}^{2})}{75 \; y_{1}^{3}
\;N}\\[15 pt]
{\cal M}^{+ \; -, \; -2}_{\gamma \; \gamma' \rightarrow G \;
(q \; \bar{q})_{PS}} &=& \; \; \; 64 \; \pi \;
\frac{D_{2}\;(z^{2}+y_{1}^{2})}{15 \; y_{1}^{3} \;N}
\end{eqnarray}
\hspace*{1cm} In the above formulas, coupling constants, as well as
charge
and color factors, were left aside. \\ [10 pt]
\hspace*{1cm} For $ \Phi_{\pi} $ we here choose two different
expressions
given in the
literature, namely  that of Chernyak and Zhitnitsky {\bf [12]}
\begin{eqnarray}
\Phi_{\pi}^{CZ}(x) &=& 5 \; \sqrt{3} \; f_{\pi} \; x \; (1-x) \; (2 x
- 1)^{2}
\end{eqnarray}
and the so-called asymptotic distribution amplitude {\bf [11]}
\begin{eqnarray}
\Phi_{\pi}^{as}(x) &=& \sqrt{3} \; f_{\pi} \; x \; (1-x)
\end{eqnarray}
where $ f_{\pi} $ is the pion leptonic decay constant. \\[10 pt]
\hspace*{1cm} Convolution of $ \Phi_{\pi} $ with formulas (5) - -
(24) then
leads us to the amplitudes
$ {\cal M}^{\lambda \; \lambda', \; \Lambda}_{\gamma \;
\gamma' \rightarrow G \; \pi^{0}} $ needed. \\
\hspace*{1cm} Therefrom one obtains the corresponding differential
cross
section :
\begin{eqnarray}
\frac{d \sigma_{\gamma \; \gamma' \rightarrow G \; \pi^{0}}}{dt} & =
& 4 \;
\frac{ \pi^{3} \alpha^{2} \alpha^{2}_{s}}{s^{2}} \; f^{2}_{ch} \;
f^{2}_{c} \sum_{ \lambda \; \lambda', \; \Lambda} \;
\left|{\cal M}^{\lambda \; \lambda', \; \Lambda}_{\gamma \;
\gamma' \rightarrow G \; \pi^{0}}\right|^{2}
\end{eqnarray}
where the charge and color factors are easily obtained as :
$ f^{2}_{ch} = 1/18 $, $ f^{2}_{c} = 2/3 $.\\[10 pt]
\hspace*{1cm} This expression of the cross section still contains an
undetermined constant,
$ f^{2}_{L} $ (see formula (1)). In order to eliminate it, we use the
same
procedure as in Refs. {\bf [1]} and {\bf [2]}, i.e. we write :
\begin{eqnarray}
\frac{d \sigma_{\gamma \; \gamma' \rightarrow G \; \pi^{0}}}{dt} \;
B(\; G \; \rightarrow \; x \; y) & \equiv &
\frac{d \sigma_{\gamma \; \gamma' \rightarrow G \; \pi^{0}}}{dt} \;
\frac{ \Gamma( J/\Psi \rightarrow G \; \gamma) \;
B(\; G \; \rightarrow \; x \; y)}{\Gamma( J/\Psi \rightarrow G \;
\gamma)}
\end{eqnarray}
where $ B(\; G \; \rightarrow \; x \; y) $ is the branching ratio for
glueball
decay in a given channel (actually we shall only consider the main
decay
channel for each glueball candidate). Then the numerator of the
second factor
on the r.h. side of (28) is given by experimental measurements, while
for its
denominator we use the expression computed by Kada et al. {\bf
[1,13]}.
In this way we
get rid, actually, not only of $ f^{2}_{L} $, but also of the (not
very well
defined) factor $ \alpha^{2}_{s} $. Eventually no free parameter is
left.
\\[10pt]
\hspace*{1cm} Our results are shown in Figs. 3, 4 and 5, for the
three glueball
candidates $ \eta (1440) \; ( J^{PC}=0^{-+}) $,
$ f_{2} (1720) \; ( J^{PC}=2^{++})$ and
$ X (2220) \; ( J^{PC}=0^{++} \; \mbox{or} \; 2^{++} \; \mbox{or} \;
4^{++})$
and with the same assumptions
 as in Ref. {\bf [1]} regarding the values of $ L, \; S $ {\bf [14]}.
In Fig. 3 we
also show, for comparison, the prediction of the model of Ref. {\bf
[11]} for
$ \gamma \; \gamma' \rightarrow \pi^{0} \; \pi^{0} $.
\section{Discussion}
\hspace*{1cm} Our results call for several comments : \\
\hspace*{1cm} {\bf (i)} ~ Quite generally, the yields
obtained with the Chernyak-Zhitnitsky pion distribution amplitude are
slightly
higher (by a factor of 3-4) than those provided by the asymptotic
one, while
the shape of the curves is very similar. It should be recalled that
the
Chernyak-Zhitnitsky distribution function should be the more reliable
one,
since it allows for a correct normalization of the charged pion's
electromagnetic form factor with a realistic value of $ \alpha_{s} $.
\\[10 pt]
\hspace*{1cm} {\bf (ii)} ~The validity of the QCD perturbation
expansion
($ p_{T} \stackrel{>}{\sim} 1 $ GeV ) entails a lower limit
$ \sqrt{s_{min}} \approx 2 \; M $ for the $ \gamma \; \gamma $ c.m.
energy
range where our model may be checked. With such a limit, our
relativistic
approximation $ (M/\sqrt{s} \rightarrow 0) $ should be  justified as
well, all
the more as it can be shown that, if we make a series expansion of
the cross
section in powers of $ M^{2}/s $, only even powers occur in that \\
expansion ;
i. e. the first term here neglected is of order $ M^{2}/s $. \\[10
pt]
\hspace*{1cm} {\bf (iii)} ~Cross sections for
$ \gamma \; \gamma \rightarrow G \; \eta $ and
$ \gamma \; \gamma \rightarrow G \; \eta' $ are easily derived from
those here
obtained if one assumes that the wave functions of $ \pi, \; \eta, \;
\eta' $
are of the same type. Then, taking the experimentally determined
values
{\bf [15]} $ f_{\eta} \simeq 0.98 f_{\pi} $, $  f_{\eta'} \simeq 0.84
f_{\pi} $,
 and $ \theta_{\mbox{mixing}} \simeq -~23^{\circ} $, one obtains for
the
relative yields of $ G~\pi^{0}~:~G~\eta~:~G~\eta' $ production the
approximate values 1 : 1.4 : 1.1.\\[10 pt]
\hspace*{1cm} {\bf (iv)} ~In a recent paper, Wakely and Carlson {\bf
[16]}
computed the same cross section, using a different procedure, namely
applying
the model of Ref. {\bf [11]} for both the pion and the glueball, with
the
glueball distribution amplitude $ \Phi_{G} $ determined by QCD sum
rules. They
gave numerical results for the production, in a $ \gamma \; \gamma $
collision,
 of an assumed pseudoscalar glueball of mass $ 2 $ GeV together with
a
$ \pi^{0} $.
 Thus in their work and ours different glueball candidates involving
different
masses and/or quantum states are considered. Therefore, strictly
speaking, it
would not make much sense to compare their results with ours.
Nevertheless,
one may observe that they are in the same ballpark. \\ [10 pt]
\hspace*{1cm} {\bf (v)} ~For the $ \eta(1440) $, for the quantum
states
``$ L=m $'' and ``$ L=2 $'' (see {\bf [14]}) of the  $ f_{2}(1720) $
and for
the quantum state $ J=4 $ of the $ X(2220) $, our predictions based
on the
Chernyak-Zhitnitsky distribution amplitude lead to values of
$ \sigma ( e^{+} \; e^{-} \rightarrow e^{+} \; e^{-} \; G \;
\pi^{0})$
(integrated above $p_{T} \simeq 1$ GeV) that should be
of the order of $10^{-37} {\rm{cm}}^{2}$ at
$ \sqrt{s} = 200 $ GeV (LEP II energy). \\ [10 pt]
\hspace*{1cm} {\bf (vi)} The formalism here used can be extended, in
a trivial
way, to the computation of glueball plus meson production in hadron
collisions.

\begin{center}
{\bf Acknowledgments}
\end{center}
\hspace*{1cm} The authors are grateful to Profs. M. Froissart and P.
Kessler
for constructive remarks and careful reading of the manuscript.
\newpage
\section{References and Notes}
{\bf [1]} E. H. Kada, P. Kessler, and J. Parisi : Phys. Rev. D 39
(1989)
2657.\\
See also : The same : Proc. IXth Int. Workshop on Photon-Photon
Collisions,
San Diego 1992, eds. D. O. Caldwell and H. P. Paar (World Scientific
Singapore
, 1992), p. 283.\\[10pt]
{\bf [2]} L. Houra-Yaou, P. Kessler, and J. Parisi : Phys. Rev. D 45
(1992) 794.\\[10 pt]
{\bf [3]} D. L. Scharre et al., Mark II Coll. : Phys. Lett. 97 B
 (1980) 329.\\
C. Edwards et al., Crystal Ball Coll. : Phys. Rev. Lett. 49
(1982) 259.\\
J.E. Agustin et al., DM2 Coll. : LAL/85-27 (1985).\\
J. Becker et al., Mark III Coll. : SLAC-PUB-4225 (1987).\\[10pt]
{\bf [4]} Z. Bai et al., Mark III Coll. : Phys. Rev. Lett. 65
 (1990) 2502; \\
T. Bolton et al., Mark III Coll. : SLAC-PUB-5632 (1991).\\
J.E. Augustin et al., DM2 Coll. : LAL/90-53 (1990).\\[10pt]
{\bf [5]} C. Edwards et al., Crystal Ball Coll. : Phys. Rev. Lett. 48
(1982) 458. \\
R. M. Baltrusaitis et al., Mark III Coll. : Phys. D 35
 (1987) 2077.\\
J. E. Augustin et al.. DM2 Coll. : Phys. Rev. Lett. 60
 (1980) 223.\\[10pt]
{\bf [6]} T. A. Armstrong et al., WA 76 Coll. : Z. Phys. C 51
 (1981) 351.\\[10pt]
{\bf [7]} L. P. Chen, Mark III Coll. : Nucl. Phys. B (Proc.
Suppl.) 21 (1991) 149 ; SLAC-PUB-5669 (1991); Ph. D. Thesis,
SLAC-Report 386 (1991).\\[10pt]
[8] G. Eigen : Proc. IXth Int. Workshop on Photon-Photon Collisions,
San Diego 1992, eds. D. O. Caldwell and H. P. Paar (World Scientific
Singapore,
 1992), p. 291.
A. Palano : same Proc., p. 308.\\[10pt]
[9] K. Einsweiler, Mark III Coll. : Ph. D. Thesis, SLAC-Report
272 (1984) ; R. M. Baltrusaitis et al., Mark III Coll. :
SLAC-PUB-3786
(1985).\\[10pt]
[10] B. V. Bolonkin et al., ITEP Coll. : Nucl. Phys. B309
 (1988) 426.\\
D. Aston et al., LASS Coll. : Phys Lett. 215B (1988) 200 ;
Nucl. Phys. B301 (1988) 525.\\[10pt]
[11] G. P. Lepage and S. J. Brodsky, Phys. Rev. D 22 (1980) 2157.
S. J. Brodsky and G. P. Lepage, ibid. 24 (1981) 1808 ; 24
 (1981) 2948. \\[10 pt]
[12] V. L. Chernyak and A, R, Zhitnitsky, Nucl. Phys. B 201
 (1982) 492.\\[10 pt]
[13] Actually we have systematically multiplied those decay widths by
a
factor of 4, because in the first paper of Ref. [1] the helicity
amplitudes
of the process
$ J/\Psi \rightarrow \gamma \; G $ had been underestimated by a
factor of 2. On
the other hand, we have corrected two misprints that appeared in that
paper,
regarding the widths of $ J/\Psi $ radiative decay into glueballs
with quantum numbers $ J=0, \; L=S=1 $ and $ J=4, \; L=S=2 $.\\[10
pt]
[14] For $ f_{2}(1270) $, the state called ``$ L=m $ '' in Ref. [1]
is a mixture of states $ L=0,\; S=2 \; \mbox{and} \; L=2, \; S=0 $
with their
respective weight coefficients $ A_{02} \; \mbox{and} \; A_{20} $
connected by
the relation $ A_{20}\; R''_{2}(0)/[A_{02} \; M^{2} \; R_{0}(0)] =
0.27 $ ; the
state called ``$ L=2 $'' is a mixture of states $ L=2,\; S=0 $ and $
L=2, \;
S=2 $ with their respective weight coefficients related by
$ A_{20}/A_{22} = -6.5 $. Both mixtures have been adjusted in such a
way that
they fit the experimental ratios of helicity amplitudes measured in
the process
$ J/\Psi \rightarrow \gamma f_{2}(1720) $. \\[10 pt]
[15] H. Aihira et al., TPC/Two-Gamma Coll. : Phys. Rev. Lett.
64, 172 (1990).
\\[10 pt]
[16] A. B. Wakely and C. E. Carlson : Phys. Rev. D 45 (1992) 1796.
\section{Figure Captions}
{\bf Fig. 1.} ~~ Kinematic schemes \\
(a) for the process $ \gamma \; \gamma' \rightarrow \pi^{0} \; G $ in
the
center-of-mass frame of $ \gamma $ and $ \gamma'$ (frame A), \\
(b) for the process $ \gamma \; \gamma' \rightarrow \pi^{0} \; g_{1}
\; g_{2} $
 ~in the center-of-mass frame ~of ~$ g_{1} $ and $ g_{2} $~~(frame
B). \\[10 pt]
{\bf Fig. 2.} ~~ Feynman graphs for the subprocess
$ \gamma \; \gamma' \rightarrow q \; \bar{q} \; g_{1} \; g_{2} $.
Others graphs,
 providing the same contribution to the helicity amplitudes of the
process\\
  $ \gamma \; \gamma' \rightarrow \pi^{0} \; G $, are derived
therefrom by
exchanging $ g_{1} $ and $ g_{2} $. \\[10 pt]
{\bf Fig. 3.} ~~ Predictions of the differential cross section for
$ \gamma \; \gamma' \rightarrow \eta(1440) \; \pi^{0} $, with the
Chernyak-Zhitnitsky pion distribution amplitude (solid line) and the
asymptotic
one (dashed line). For comparison, we also show the predictions of
the
Brodsky-Lepage model for $ \gamma \; \gamma' \rightarrow \pi^{0} \;
\pi^{0} $
with the Chernyak-Zhitnitsky pion distribution amplitude (dash-dotted
line)
and the asymptotic one (dotted line). \\[10 pt]
{\bf Fig. 4.} ~~ Predictions of the differential cross section for
$ \gamma \; \gamma' \rightarrow f_{2}(1720) \; \pi^{0} $ with the
Chernyak-Zhitnitsky and the asymptotic pion distribution amplitude,
considering various quantum states of the $ f_{2}(1720) \; : \; L=0 $
(solid line) ; `` $ L=2 $ '' (dotted line) ; `` $ L=m $ '' (dashed
line)
(see {\bf [1, 2]}). \\[10 pt]
{\bf Fig. 5.} ~~ Predictions of the differential cross section for
$ \gamma \; \gamma' \rightarrow X(2220) \; \pi^{0} $ with the
Chernyak-Zhitnitsky and the asymptotic pion distribution amplitude,
considering various quantum states of the $ X(2220) \; : \; J=L=0 $
(solid line) ; $  J=2, \; L=0  $ (dashed line) ; $  J=4, \; L=2  $
(dotted line). \\[10 pt]
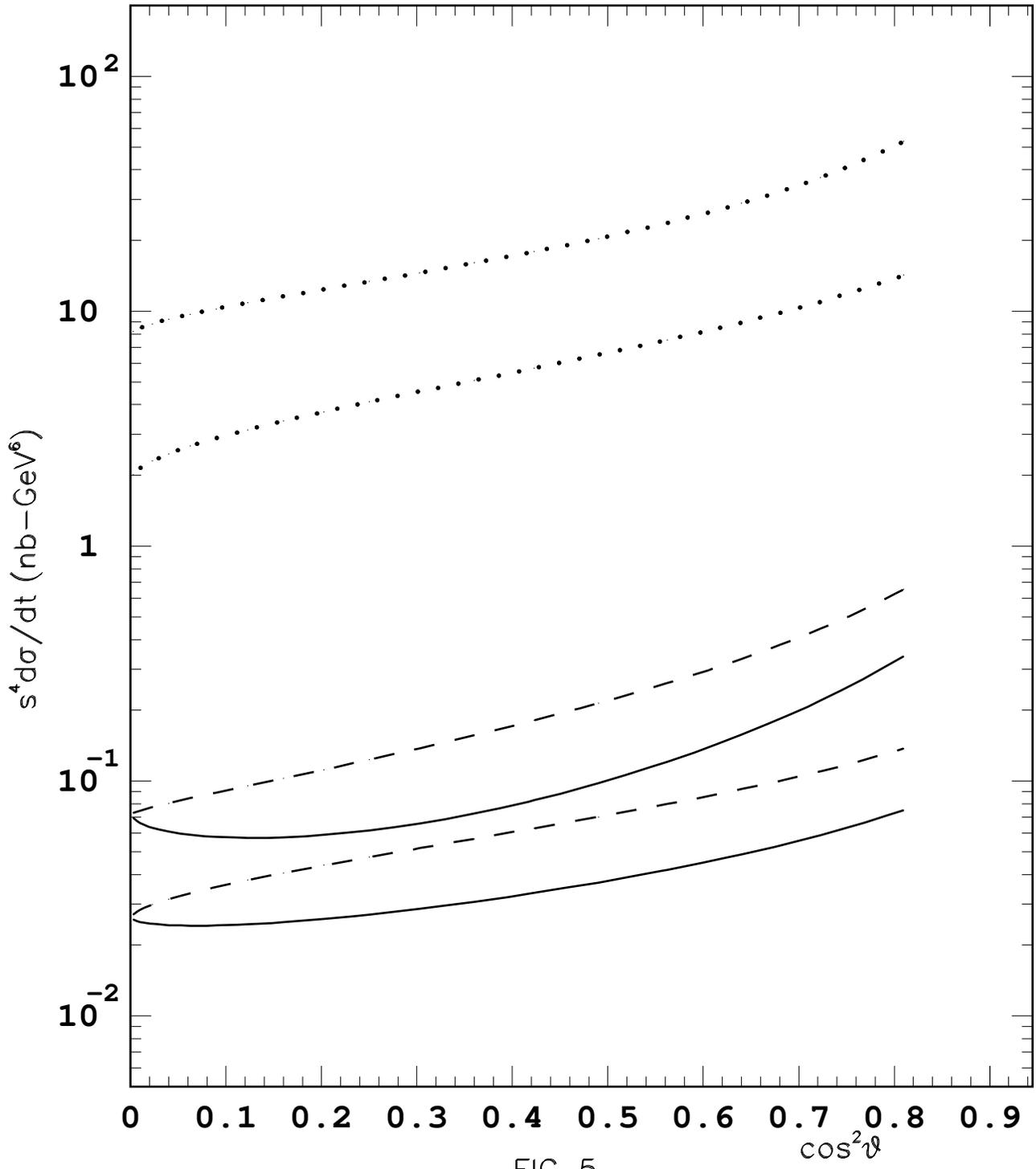
\end{document}